\documentclass[12pt,preprint]{aastex}

\newcommand{\beq}{\begin{equation}}
\newcommand{\eeq}{\end{equation}}

\newcommand{\gsim}{\gtrsim}

\begin{document}

\title{The H$\alpha$ Luminosity Function of Morphologically Classified
Galaxies in the Sloan Digital Sky Survey}

\author{Osamu Nakamura\altaffilmark{1},
Masataka Fukugita\altaffilmark{1},
Jon Brinkmann\altaffilmark{2},
Donald P. Schneider\altaffilmark{3}}
\altaffiltext{1}{Institute for Cosmic Ray Research, University of Tokyo,
Kashiwa 277 8582, Japan}
%%\email{osamu@icrr.u-tokyo.ac.jp}
\altaffiltext{2}{Apache Point Observatory, Sunspot, NM 88349, U. S. A.}
\altaffiltext{3}{Department of Astronomy and Astrophysics, Pennsylvania 
State University, University Park, PA 16802, U. S. A.}

\begin{abstract}

We present a study of the 
H$\alpha$ line emission from a sample of 1482 optically-selected,
morphologically-classified bright galaxies (median redshift of 0.05)
derived from the Sloan Digital Sky Survey. The luminosity function is 
calculated for each morphological class and for the total sample.
The luminosity function fitted with the Schechter form gives a slope 
$\alpha=-1.43\pm 0.10$ for the total sample and the H$\alpha$ luminosity 
density is $10^{39.31\pm 0.04{+0.10\atop-0.07}}h$ erg s$^{-1}$Mpc$^{-3}$,
where the first error is statistical and the second is systematic. 
% where $h$ is the Hubble constant normalised to 100 km s$^{-1}$Mpc$^{-1}$. 
This value is consistent with that derived by
Gallego et al. (1995), but this agreement is caused by a fortuitous
cancellation of their neglect of stellar absorption that affects 
the estimate of extinction corrections and a significant 
sample incompleteness of emission line galaxies. 
The fraction of H$\alpha$ emitters monotonically increases from 
early (a few \% for ellipticals) to late types 
(100\% for irregular galaxies), whereas strong emitters 
exist in all classes of morphological types. We find that
83\% of the luminosity density 
comes from spiral galaxies, 5\% from irregular galaxies, and  
9\% from early type galaxies; a small number of morphologically disturbed 
galaxies contribute by 3\%.

\end{abstract}

\section{Introduction}

Much attention has been devoted to the global star formation rate of galaxies as 
a function of redshift. These studies have advanced our understanding of
how galaxies
evolved towards the present epoch (Gallego et al. 1995; Madau et al. 1996;
Lilly et al. 1996; Treyer et al. 1998; Tresse \& Maddox 1998; 
Steidel et al. 1999;  Glazebrook et al. 1999; Sullivan et al.
2000; Tresse et al. 2002). 
Despite its obvious importance, the estimate of the star formation rate 
at zero redshift, which is the boundary value for the star 
formation history, still largely relies on the pioneering work of 
Gallego et al. (1995) which was based on Schmidt objective prism plates
searching for strong H$\alpha$ emission galaxies with observed
equivalent widths 
EW(H$\alpha$+[NII])$>$10\AA.
The evolution of galaxies is inferred by comparing the global star 
formation rates at higher redshift 
with Gallego et al.'s value at $z\approx0$. Radio investigations have  
suggested star formation rates higher than Gallego et al.'s by a factor of two
(Cram 1998; Serjeant, Gruppioni \& Oliver 2002). 
The follow-up work for Gallego et al.'s sample by Gil de Paz et al. 
(2000) %and Alonso-Herrero (1996) 
gave valuable information as to the nature of star forming galaxies, 
yet the question of when galaxies 
show star formation activity among their entire population  
remains uncovered. Also a poorly understood aspect is the 
star formation activity as a function of 
morphology of galaxies, which constitutes another
dimension of galaxy formation. 

In fact, there are a few pieces of work which studied the star formation rate
as a function of morphologies of galaxies (Kennicutt \& Kent 1983;
Ryder \& Dopita 1994; hereinafter RD94). These studies, however, were
based on nearby galaxy samples, which were preselected by choice,
while ideally one want to study homogeneous galaxy samples. 
The primary difficulty of this approach for nearby galaxies 
is to obtain 
accurate spectroscopic information for a homogeneous galaxy sample 
over a large area of the sky. 
The Sloan Digital Sky Survey (SDSS; York et al. 2000) 
overcome this difficulty with its spectroscopic data for 
a large number of homogeneously selected galaxies. The SDSS data also allow
significantly more accurate treatments as to the control over selection 
effects, such as those expected in line-selected samples,
and systematic errors arising from stellar absorption, extinction, 
and contamination from AGN.

In this paper we use the optically-selected and morphology-classified
bright galaxy sample from the Sloan Digital Sky Survey 
for the northern equatorial stripe.
The sample is flux-limited with $r^*\leq 15.9$ mag after 
dereddening corrections,
and morphological classification is given by visual inspections.
This sample, with median redshift $z=0.05$, has been used to derive
luminosity functions (LF) of morphologically selected galaxies 
(Nakamura et al. 2003; hereinafter N03). 
The SDSS spectroscopy covers from 3900 to 9000 \AA~with the resolution
of 2-5 \AA. Redshifts are identified and EWs are measured with 
the spectroscopic pipeline.
We use H$\alpha$ as an indicator of star formation activity.
The SDSS spectroscopy is sufficient to resolve H$\alpha$ and 
[N II]$\lambda\lambda$6548,6583 doublet. The advantage of using H$\alpha$
as the star formation indicator is that this emission 
arises from the ionising flux that is produced by massive short-lived 
stars compared, for instance,
to the indicators using the UV light at 1500-2800\AA~which survives, 
or even increases, for a significant duration of time after stars 
formed (e.g., Tresse \& Maddox
1998; Glazebrook et al. 1999). Another advantage is that the 
extinction correction is smaller and readily corrected for by using
the Balmer line ratio. This contrasts to UV emission, which  
suffers from large extinction but also an uncertainty in
extinction laws (Calzetti et al. 1994). The most important disadvantage
with the SDSS spectroscopy for our purpose is that it uses
fibres of small apertures, and the aperture correction is essential.
We circumvent this problem by using the empirical radial profile of  
H$\alpha$ obtained by RD94 to calculate the total
H$\alpha$ flux.

Our prime purpose is to investigate the H$\alpha$ 
emissivity distribution for each class of morphological type of galaxies,
and its contribution to global H$\alpha$ luminosity in the nearby 
universe ($z\leq0.12$) 
using a well-defined optically flux-limited sample.
We prefer to use traditional morphological classification obtained 
by visual inspections, since spectroscopic or colour
indicators are too sensitive to small amounts of star formation activity, 
whereas visual morphology reflects dynamical evolution of
galaxies more faithfully. The drawback of this approach  
is the small size of the sample 
(1600 galaxies).
Notwithstanding, we emphasise that the present study is based on a
galaxy sample with homogeneous morphological classification and 
accurate photometry; the resulting H$\alpha$ emission sample is
significantly larger than those used in previous studies.

% We also spend a significant amount of space to discuss the 
% H$\alpha$ LF and the H$\alpha$ luminosity density. A detailed comparison
% is given with the work of Gallego et al. (1995), and discuss how robust
% is their results, especially the H$\alpha$ luminosity density.

In section 2 we describe the data and various corrections that are crucial
to our study. The H$\alpha$ LF is derived in section 3 for each morphological
class. A detailed comparison is given with the work of Gallego et al. (1995)
for the total H$\alpha$ luminosity density in this section. 
Morphological-type and colour dependence of H$\alpha$ emissivity is 
presented in section 4, and implications of our results are discussed in
section 5.

We adopt the cosmology $\Omega=0.3$ and $\lambda=0.7$ throughout this
paper, although the physical parameters of our objects, being at such 
low redshift, depend little on $\Omega_0$ and $\lambda_0$.
We use the standard notation $h=H_0/100$km s$^{-1}$Mpc$^{-1}$
to demonstrate the dependence of quantities upon $H_0$.
% but use $H_0=70$ km s$^{-1}$Mpc$^{-1}$ when a specific
% value is needed.

\section{The emission line galaxy sample}

\subsection{Data}

Our sample is derived from the photometric images (Gunn et al. 1998;
Hogg et al. 2001; Pier et al. 2003) of Run 752/756 of 
early SDSS observations [the area lies along the celestial equator
for $9^h40^m36^s\le \alpha(J2000)\le 15^h43^m53^s$].
The observations cover 229.7 deg$^2$ in the five colour band
photometric system (Fukugita et al. 1996), calibrated by
the standard star network established by the work at USNO
(Smith et al. 2002). The bulk of data were published
in Early Data Release (Stoughton et al. 2002), and are also included in
Data Release One (DR1, Abazajian et al. 2003). Spectroscopic
information is obtained for galaxies that satisfy accurately
defined selection criteria (Strauss et al. 2002; Blanton et al. 2003b). 
The morphological classification was performed by visual inspections
for all galaxies with dereddened Petrosian magnitude 
of ${r^{\ast}}_P\leq 15.9$ included in the rectangular area
of Run 752/756\footnote{
Although new photometry is available in DR1, we use old photometry of
EDR since the morphologically classified catalogue is made using the
sample based on old photometry and the completeness would be lost
in faint magnitudes if new photometry is adopted.}. 
The morphological index $T=0-6$ is assigned to
E, S0, Sa, Sb, Sc, Sd and Im. The galaxies that show largely disturbed
morphologies are given $T=-1$.
The size of galaxy sample with
spectroscopic information is 1600 (See Table 1 of N03).

Among 1600 galaxies, we drop 21 objects because of either low ($<85$\%)
redshift confidence or misallocation of fibres to overlapping stars,
29 galaxies because of poor photometry due to deblending of 
neighboring objects, and 59 galaxies that fall either
$z<0.01$ or $z>0.12$. We remove an additional
9 galaxies with ${r^{\ast}}_P<13.2$,
at which magnitude the spectroscopic survey completeness is 
lowered due to the SDSS target selection algorithms. 
%An additional one galaxy is dropped due to the fact that the fibre is centred 
%on a HII region off the centre of galaxy. 
These cuts leave 1482 galaxies
for our consideration. This data set is the same 
%(except for the one galaxy mentioned in the preceding sentence) 
as the sample from which
the morphologically-dependent luminosity function was 
derived in N03 (sample 5 in Table 1 of N03), and is recapitulated in
Table 1 below. These
selections do not introduce any significant bias into the sample
with respect to morphology or luminosity.

%      1600
% Fs   1579   flag_sp=good                                         SAMPLE-A
% Fp   1561   flag_ph=good
% Fps  1550   flag_sp=good, flag_ph=good
%      1534   flag_sp=good, flag_ph=good,        z<=0.12
%      1507   flag_sp=good, flag_ph=good,  0.01<=z
%      1491   flag_sp=good, flag_ph=good,  0.01<=z<=0.12
%      1491   flag_sp=good, flag_ph=good,  0.01<=z<=0.12, r>=13.2  SAMPLE-B

The SDSS spectroscopy was carried out using fibres of $3''$ diameter
placed on the centre of each object with F/5 optics. 
The 2-5\AA~ spectral resolution
of the double spectrographs resolves H$\alpha$ emission
from [NII]$\lambda$6548 and [NII]$\lambda$6583. This contrasts to
earlier works where H$\alpha$ is often blended with nitrogen lines 
(e,g., Sullivan et al. 2000; Tresse \& Maddox 1998). 
Redshifts and EWs are automatically 
measured by the SDSS spectroscopic pipeline 
and are catalogued. We use redshifts converted into the values
in the Galactic standard of rest according to de Vaucouleurs et al. (1991).
There is little confusion in the measurement of 
H$\alpha$ equivalent widths: visual inspection 
of spectra shows that emission lines with greater
than $\approx$0.5\AA~in the observed frame 
are accurately measured for the majority of galaxies. 
Manual remeasurements of EWs for
selected galaxies using IRAF show that the
error is less than $0.5$\AA~for small EWs and less than 
10\% for larger EWs. The error mostly arises from the fit to
the continuum.  
We shall work with the rest-frame equivalent width (hereafter we always 
refer to the rest-frame EW unless otherwise noted) with measurements 
at more than 2.5$\sigma$. A total of 30 galaxies have EW$\geq 1$\AA~yet
the detection confidence is less than 2.5$\sigma$: those galaxies are
omitted from our main sample. 
We find that most cases this occurs when the H$\alpha$ line is too broad 
and/or emission of [NII]
is much stronger than that of H$\alpha$, and automated deblending of the 
H$\alpha$ lines is not properly executed. 
So, those that are missed by 2.5$\sigma$
criterion for EW$>1$\AA~are AGN like galaxies. Finally, we dropped
one galaxy for which the fibre is centred on a HII region off the 
centre of galaxy. Thus, our emission line galaxy sample contains 665 galaxies.

We show in Figure 1 the EW distribution of H$\alpha$ for
morphologically-classified sample, from the latest type (top panel) 
to earlier types. The bottom panel is the total sample. 
A negative EW signifies absorption. The solid lines correspond to
our main sample with EW$>1$\AA~ and the detection at
$>2.5\sigma$, and dotted are
galaxies that do not satisfy these criteria. Note that the ordinate is
a logarithmic. The mean of emission line strengths for
each morphological class is indicated by the diamonds. In this
average (indicated by dotted histogram), we include galaxies with 
EW$<1$\AA~and with lower confidence detection, as well as those that do not
show emission at all.
AGN-like galaxies, as discussed in Sect. \ref{sec:AGN}, 
are denoted with shading. 
The numbers are presented in Table~1, where emission line galaxies are listed
separately for non-AGN and AGN galaxies.
The bottom panel shows that for EW$>5$\AA~the galaxies that are
excluded from the sample by 2.5$\sigma$ detection are only a few,
and the sample is virtually complete. The sample incompleteness for
EW$=1-5$\AA~ is about 14\%. We also try to include those galaxies
which do not satisfy the $>1$\AA~ or the 2.5$\sigma$ criterion in 
our analysis and examine how the results change.

\subsection{Stellar Absorption Corrections}

The measured Balmer emission line equivalent widths must be corrected for
stellar absorption. The absorption 
not only decreases H$\alpha$ emission flux,
but also significantly affects estimates of reddening corrections
inferred from Balmer line ratios.
Figure 2 shows the relation between  EW(H$\alpha$) and
EW(H$\beta$)$\times F_c({\rm H}\beta)/F_c({\rm H}\alpha)$,
where $F_c$ is the continuum flux densities at the two lines 
[$F_c({\rm H}\beta)/F_c({\rm H}\alpha)\simeq1.0$ 
for galaxies with EW(H$\alpha$)$\le25$\AA~ in our sample].
The relation plotted in this figure is equivalent to
that between the line fluxes, $F($H$\alpha$) and $F($H$\beta$).
Considering galaxies with EW(H$\alpha$)$>$10.0{\AA}, for which
corresponding EW(H$\beta$) is mostly larger than 1.0{\AA} 
and hence the H$\beta$
measurement is reliable, we obtain with a $\chi^2$ fit,
\begin{equation}
{\rm EW(H}\beta)\times \frac{F_c({\rm H}\beta)}{F_c({\rm H}\alpha)}
      =-1.58+0.246\times{\rm EW(H}\alpha),
\label{eq:ew}
\end{equation}
where 3$\sigma$ rejection is applied. The existence of a non-zero constant
term indicates the presence of stellar absorption. If we
take the H$\alpha$ absorption to have an equivalent width of 
$-1.8${\AA}, as indicated from
the conspicuous clustering of non-emission line galaxies seen in the
figure, we obtain from eq. (\ref{eq:ew}) 
EW(H$\beta)_{\rm abs}=-2.0$\AA, which is also
consistent with the position of EW(H$\beta)$ for non-emission line
galaxies. 

An independent check of the effect of stellar absorption is 
carried out assuming  
EW(H$\gamma)_{\rm abs}$ =EW(H$\beta)_{\rm abs}$ 
(e.g., McCall, Rybski \& Shields 1985) for 168 galaxies with
EW(H$\gamma$) detection at $>2.5\sigma$ out of 665 emission line galaxies.
After extinction corrections and assuming the standard
Balmer line ratios, as we describe in the next subsection,  
we obtain EW(H$\gamma)_{\rm abs}\approx -1.5$\AA.
(see Figure 3) independent of  
EW(H$\alpha)$.

In our following analysis we take EW$_{\rm abs}$(H$\alpha$)=$-1.8${\AA} and 
EW$_{\rm abs}$(H$\beta$)=$-2.0${\AA} for all galaxies. We note that the
estimate of extinction corrections depends little on the absolute value of
the stellar absorption in so far as we use the H$\beta$/H$\alpha$
line ratio from (\ref{eq:ew}).

\subsection{Extinction Corrections}  
\label{sec:extinc}

The extinction correction for the line flux is estimated from the 
H$\alpha$/H$\beta$ Balmer line ratio, which is written 

\begin{equation}
\frac{F({\rm H}\alpha)}{F({\rm H}\beta)}
= D~10^{-c[f({\rm H}\alpha)-f({\rm H}\beta)]},
\label{eq:ext}
\end{equation}
where $D$ is the ratio of the intrinsic fluxes emitted in the nebula,
$f(\lambda)=k(\lambda)/k({\rm H}\beta)$, and $c=0.4k({\rm H}\beta)E(B-V)$
with $k(\lambda)$ the Whitford extinction curve that satisfies
$k(\lambda)=A_{\lambda}/E(B-V)$.
We take Seaton's (1979) law for $k(\lambda)$, which gives
%$k({\rm H}\alpha)=2.49$ 
$k({\rm H}\beta)=3.68$ and $k({\rm H}\beta)/k({\rm H}\alpha)=1.48$.
We also evaluate the extinction using O'Donnell (1994) extinction law
with $R=3.1\pm0.2$ to investigate the systematic error.
We set $D=2.86$ assuming Baker-Menzel case B recombination for the
electron temperature $T_e=10^4$K and the density $n_e=100~$cm$^{-3}$ 
(Brocklehurst 1971). 
The change of the temperature by a factor
of two modifies $D$ only by 5\%; the change of $n_e$ affects 
$D$ very little. 

For galaxies with $>2.5\sigma$ detection of H$\beta$ 
[EW(H$\alpha$)$\gsim$10.0{\AA}], 
eq. (\ref{eq:ext}) is applied to each galaxy to derive $A_{{\rm H}\alpha}$.
The distribution of extinction is shown in Figure 4 after converting
it into $V$ band extinction $A_V$ using $A_V/A_{H\alpha}=
1.28$. The mean value of extinction is 
$\langle A_V\rangle=1.10$. This distribution
is consistent with that obtained by Sullivan et al. (2000) for
$\langle z\rangle=0.2$ galaxies ($\langle A_V\rangle=0.97$). 
The mean extinction also agrees with
the value derived by Kennicutt (1983). If stellar absorption is not
taken into account, the extinction correction would
become erroneously large by 70\%, as noted by Sullivan et al. (2000), and
the resulting extinction correction would also show an anticorrelation with
the H$\alpha$ equivalent width.
 For rather weak emitters, i.e., EW(H$\alpha$)$<$10.0{\AA} or
the H$\beta$ detection below $2.5\sigma$, 
we adopt $A_{{\rm H}\alpha}$=0.81
($\langle A_V\rangle =1.03)$ which is determined by 
the H$\alpha$/H$\beta$ line ratio
calculated from eq. (\ref{eq:ew}).  
%We note that the extinction thus derived depends little on
%the absolute value of stellar absorption. 

\subsection{AGNs}
\label{sec:AGN}

Since we are primarily interested in the star formation activity, 
we must remove H$\alpha$
contributions from AGN. We use the line diagnostic diagram employing
[NII]$\lambda$6583/H$\alpha$ versus [OIII]$\lambda$5007/H$\beta$
to identify AGN
(Baldwin, Phillips, \& Terlevich 1981; Veilleux \& Osterbrock 1987;
see also Kewley et al. 2001).
We require that four relevant lines be detected at more than 2.5$\sigma$.
Then, 363 galaxies out of 665 galaxies pass this criterion.
The line diagnostic we use is immune to dust extinction, since the lines
of each pair are close enough in wavelength. 

We use the criterion set by Kauffmann et al. (2003), which was applied
to the SDSS galaxy sample to select AGN:

\begin{equation}
\log(  \frac{[{\rm O III}]\lambda5007}{{\rm H}\beta} )
     \ge
\frac{0.61}{ \log(
       \frac{[{\rm N II}]\lambda6583}{{\rm H}\alpha} )
             -0.05} + 1.3,  \\
\label{eq:agn2}
\end{equation}
where H$\alpha$ and H$\beta$ should be corrected for stellar absorption. 89 
galaxies satisfy this criterion and are identified as Seyfert II
or LINER. The frequency of AGN thus identified is 6\%,
compared to 18\% in the Kauffmann et al. sample. 
The difference arises from weak H$\beta$ line emitters, for which special
effort is made by Kauffmann et al. to measure weak lines with
higher signal-to-noise ratio. In our sample about 3\% of galaxies with
H$\beta$ emission of EW$>$1\AA~are missed by our 2.5$\sigma$ criterion, but 
this fraction increases to  26\% for EW=0.5-1\AA~ and rapidly increases for
EW$<0.5$\AA~. Most of the difference in numbers of AGN between Kauffmann et 
al. and this study arises from galaxies with EW(H$\beta)<0.5$\AA, which are not
important to us. If we use two-line diagnostic with only H$\alpha$ and [NII]
[$\log([{\rm N II}]\lambda6583/{\rm H}\alpha)>-0.3$],
the number of AGN increases to 235 galaxies which is 16\% of the 
total number of galaxies. In the next section we derive the H$\alpha$
luminosity functions, the result of which only weakly depends on whether we
reject 89 AGN or 235 `AGN'.

\subsection{Aperture corrections}

Since the line flux is measured with fibres of a $3''$ aperture, aperture
corrections are essential to estimate total H$\alpha$ luminosity of 
galaxies. There is also a problem with
spectrophotometric calibration, the error of which is about
$\approx 20$\% (Stoughton et al. 2002), which we confirmed
for our sample. 
%This error can be removed by comparing the spectroscopic 
%flux with the photometric flux.

The aperture correction is made by using empirical radial 
profiles of H$\alpha$ and broad band fluxes obtained by 
RD94. These authors studied H$\alpha$ and 
$I$ band surface brightnesses for 34 nearby S0 and spiral 
galaxies. We show in Figure 5 the H$\alpha$ flux versus $i'$ band flux
($I$ mag of RD94 is translated to $i'$ system using Fukugita et al. 1995).
The dashed (dotted) lines show the trace of H$\alpha$ and $i'$ band 
fluxes of RD94 photometry: it moves left/upwards as the aperture 
increases and the circle denoting their `total' flux. Superimposed 
in the bundle of traces are SDSS spectrophotometric data of
H$\alpha$ and $i^*$-synthetic flux in the $3''$
aperture. We see that the SDSS data overlap very well with the RD94 
traces, except for several galaxies (denoted by dots) which deviates
from the SDSS data points.

We construct a composite growth curve of the ratio of
H$\alpha$ to $I$ band flux from 28 galaxies, rejecting 6 galaxies that
are outliers of the SDSS data. 
We integrate the H$\alpha$ and $I$ band fluxes from the centre of
galaxies outwards, and the integrated H$\alpha$ fluxes are evaluated against
the integrated $I$ band fluxes, with both total fluxes normalised to
unity.
We do not observe systematic trends in
growth curves that depend on morphologies of galaxies\footnote{This
does not mean that the ratio of the total H$\alpha$ to the total $I$ band
fluxes does not depend on morphology. RD94 showed that such a
ratio actually depends on morphology, whereas the growth curve
is nearly universal.}. 
To obtain the broad band Petrosian flux defined
by the SDSS, we extrapolate the measured flux assuming 
an exponential profile with the scale length determined by 
RD94 (this gives $\approx 8-20$\% correction). For the H$\alpha$
flux, there is a natural cutoff at 
$\mu_{{\rm H}\alpha}\approx 2.5\times10^{38}$ erg s$^{-1}$kpc$^{-2}$,
as shown by Kennicutt (1998),
and the limit of RD94 measurement agrees with this cutoff; the 
integration of the flux measured by RD94 gives the total H$\alpha$ flux.
The growth curve thus constructed is shown in Figure 6, together
with those of individual galaxies (the error bars show one sigma
at representative points).
The composite growth curve shows
that H$\alpha$ emission activity is somewhat more active in the outer
region of galaxies than in the central part, although for individual
galaxies the growth curves show a variety.  
Note that our $3''$ aperture corresponds to $(2.1-3.9)h^{-1}$ kpc 
at $z=0.05-0.10$.

We first renormalise the spectrophotometric flux by adjusting $i'$ band 
synthetic magnitude obtained from spectrophotometry to  the corresponding 
$3''$ aperture photometric magnitude to 
remove the calibration error of spectrophotometry, 
and then apply the growth curve
to find the total H$\alpha$ flux from its spectrophotometric flux and 
$i'$ band photometric fluxes at the $3''$ and the Petrosian apertures. 
Our calculation takes into account the
aspect ratio of the galaxy images, which gives only a small correction.

The aperture correction amounts to 2.40$\pm0.83$ mag for $z\simeq 0.015$, 
$2.14\pm0.59$ mag at $z\simeq 0.05$ and $1.56\pm0.44$ mag
at $z\simeq 0.1$ (the error stands for rms).
The total H$\alpha$ flux we estimate is shown in Figure 7. We do not
see any systematic variations of the high luminosity edge as a function
of redshift, suggesting that aperture corrections are properly done. 
Systematic errors from this aperture correction are estimated by changing
the composite growth curve by $\pm 1\sigma$. 
If we would apply aperture correction simply by scaling the 
synthetic H$\alpha$ flux with the broad band fluxes measured at
two apertures, we underestimate the total
H$\alpha$ flux  by 20$-$25\%.

\section{The H$\alpha$ luminosity function}

\subsection{Calculation}

We calculate the H$\alpha$ luminosity function (LF) 
for five morphological classes and
for the total sample. We treat E and E/S0-S0 separately, but when we
compute LF we add the two together since 
the number of H$\alpha$ emission E galaxies (Fukugita et al. 2003)
is too small to derive a
reliable LF for this type. 
The number of galaxies in each morphological type is given
in Table 1 above. We reject 89 AGN from the sample. We employ the step-wise
maximum likelihood estimator of Mobasher, Sharples, and Ellis (1993),
using the $r^{*}_{P}$ band LF calculated by the maximum likelihood 
estimator in N03. 
The step-wise H$\alpha$ LF is given by
\begin{equation}
\phi(L_{\rm H\alpha}) \Delta L_{\rm H\alpha} = \sum_{i=1}^{N(L_{H\alpha})}
                 \frac{\phi(M_{r^{*}_{P}}^{i})}{n(M_{r^{*}_{P}}^{i})},
\label{eq:calclf}
\end{equation}
where $\phi(M_{r^{*}_{P}}^{i})$ is the $r^{*}_{P}$ band LF,
$n(M_{r^{*}_{P}}^{i})$ is the number of objects 
that emit H$\alpha$ 
luminosity between $(L_{H\alpha}-\Delta L/2)$ and $(L_{H\alpha}+\Delta L/2)$ 
and whose $r'$ band luminosity
is in the $M_{r^{*}_{P}}^{i}$ magnitude bin. The function in the sum 
corresponds to the inverse of a survey volume in a given magnitude 
bin given by the maximum likelihood estimator. This estimate is
not affected by inhomogeneity of the sample, unlike the $1/V$(max)
estimator for which the function in the sum is simply 
the inverse volume of visibility. The results are given in 
Figure 8.

The H$\alpha$ LF we obtained shows that the Schechter (1976)
function 
\begin{equation}
\phi(L_{\rm H\alpha})dL_{\rm H\alpha}=\phi^*
\left(\frac{L_{\rm H\alpha}}{L^*}\right)^\alpha
\exp\left(-~\frac{L_{\rm H\alpha}}{L^*}\right)
\frac{dL_{\rm H\alpha}}{L^*}
\end{equation}
gives a reasonable
fit. We carried out a  $\chi^2$ fit, rejecting
data in the bins where the number of galaxies is one (we include such
data, however for the LF of Im type, as this entire sample is very small).
The Schechter function fit suffers a binning artefact. We, therefore,
carry out 10 fits shifting the binning by 1/10 the interval of bins,
and smear the resulting parameters.
We also removed the data in the faintest bin, which may be affected by the
binning.  
The results are presented in Figure 8 and the parameters obtained are shown in
Figure 9, where the error contours are one and two standard deviations of 
the  $\chi^2$ fit. The parameters and one standard deviation errors
are also given in Table 2. 
We attempt to estimate errors by the jackknife method dividing 
the sky region into 20 segments. We find that those errors are comparable to 
one standard deviation errors we quoted.

We calculate H$\alpha$ luminosity density for each morphological class
by integrating the
Schechter function to zero luminosity. This is compared to the 
luminosity density directly calculated as 
$\log {\cal L}({\rm H}\alpha)=\sum_i L_i/V_i^{\rm max}({\rm eff})$
where $V_i^{\rm max}({\rm eff})=
n(M_{r^{*}_{P}}^{i})/\phi(M_{r^{*}_{P}}^{i})$ that appears in
eq. (\ref{eq:calclf}). The two estimates show good agreement for
late types, but the two differ by 0.1 dex for
E-S0 and Im, for which statistics are very small and the
Schechter function fits are rather poor: 
for these cases the direct sum is probably more reliable, and the
luminosity densities from the direct sum are adopted for our
discussion. 
The total luminosity density from the Schechter function is 
$\log {\cal L}({\rm H}\alpha)({\rm erg~s}^{-1}{\rm Mpc}^{-3})=(39.31\pm0.04)+\log h$,
which agrees with $39.33+\log h$ from the direct sum.
The sum of the component luminosity densities amounts to 98\% of the total
(when we discuss the composition of the luminosity densities we 
renormalise the component sum to 100\%).
Note that the $r'$ band luminosity density of N03 is 
$2.00\times10^8L_\odot h$(Mpc)$^{-3}$.

We note that the incompleteness of spectroscopic and/or photometric
samples is already corrected by the use of $r'$ band LF. The final
correction should account for the under or overdensity of the   
bright galaxies in the northern equatorial stripes. 
The $r'$ band luminosity density of N03 has virtually no offset
against the global value of Blanton et al. (2003a)\footnote{
%We estimate that the sample is underdense by 3.5\% taking 
%${\cal L}_{r^*}=2.07\times 10^8hL_\odot$Mpc$^{-3}$ as 
%the reference
In N03 we noted that
the luminosity density of bright ($r^*<15.9$) galaxy sample is
lower than the total spectroscopic sample with $r^*<17.88$ 
(Blanton et al. 2001) by 29\%. A further study (Blanton et al. 2003a), 
however, showed that what is {\it overdense} is 
the spectroscopic sample of the northern equatorial   
strip compared to the global value.
In the $r'$ pass band, the global luminosity density is approximately 
$2.0\times10^8L_\odot h$(Mpc)$^{-3}$.}.   
%With this correction 
%$\log {\cal L}({\rm H}\alpha)({\rm erg~s}^{-1}{\rm Mpc}^{-3})=39.33\pm 0.04$ 
%at $h=1$.

Before discussing the results we address the problem of systematic 
uncertainties.
The first issue is the effect of the threshold set for H$\alpha$
detection (EW(H$\alpha)>$1\AA). In order to test the effect we  decrease the
detection threshold to EW(H$\alpha$)$\ge-0.8$({\AA}) removing the
2.5$\sigma$ detection criterion. 
%Since we use EW$_{\rm abs}$(H$\alpha$)=$-1.8${\AA},
The new threshold corresponds to selecting galaxies with
EW(H$\alpha)_{\rm em}\ge1.0$({\AA}) after absorption correction.
%The results are displayed in Figure 8 (a). 
With this change 
the numbers of galaxies increase to
51, 158, 394, 337, and 10 for E, S0, early spiral, late
spiral, and Im galaxies, respectively (for weak line emitters, 
we do not reject galaxies
that possibly satisfy the AGN criterion).
While this change increases the number of emission line galaxies by 50\%, 
it does not appreciably affect the H$\alpha$ LF
for the late-type and the total samples.
The change is observed in the fainter bins for E
and S0 galaxies, which are dominated by non or weak
H$\alpha$ emitters around the threshold. The brighter part of the E plus S0
luminosity function is unchanged. The
effect on the H$\alpha$ luminosity density is only +4\%. 

There is some uncertainty in the estimate of the stellar absorption
correction. A $\pm0.5$\AA~uncertainty in the stellar absorption strength
yields a $-$3 to +7\% change in the luminosity density.
A complete neglect of stellar absorption would affect the H$\beta$/H$\alpha$
flux ratio for weak emitters, and leads to a significant overestimate 
of extinction corrections: this results in the H$\alpha$ luminosity
about 0.4 dex brighter, and the luminosity density 
50\% larger than the true value (NB: $\phi^*$ decreases). 
The change of extinction law from the
Seaton (1979) to the O'Donnell (1994) relations 
(with $R=3.1$) does not significantly affect the results:
the luminosity density increases by only +2\%. Allowing for 
$R=3.1\pm0.2$, the uncertainty is $-$2\% to +10\%. We also allocate
an uncertainty of $\pm 10$\% from the error for the estimate of
$\langle A_{H\alpha}\rangle$. 

If we exclude additional 146 weak H$\beta$ `AGN' as discussed in 
sect. \ref{sec:AGN}, the luminosity density decreases by 5\%.
The shape of the H$\alpha$ LF changes very little. A complete inclusion of
all AGN would increase the H$\alpha$ luminosity density, but it
is no more than by +18\%.

The systematic uncertainties we considered are summarised in Table 3,
expressed in the form of the change in the luminosity density.
We also include the cases which, we believe, are unrealistic
in order to show the effect if we ignore
the relevant considerations, as are occasionally seen in some  
previous works.  The systematic error is obtained by adding the entries
with asterisk in quadrature: our estimate is +0.10 dex and $-$0.07 dex
for $\cal L$.

\subsection{The results}

The characteristics of our H$\alpha$ LF are as follows.

1. The H$\alpha$ LF for the total sample is fitted well by a Schechter
   function with $\alpha=-1.43\pm 0.10$ and $\log L^*=41.68\pm 0.10$. 
   % The faint end slope and the characteristic luminosity
   % are consistent with those of Gallego et al. (1995) for their 
   % $z<0.045$ sample investigated by
   % Gallego et al. (1995). The normalisation, however, is slightly 
   % lower. 
   The H$\alpha$ luminosity density derived from our LF,
\begin{equation} 
   \log {\cal L}({\rm H}\alpha)({\rm erg~s^{-1}Mpc^{-3}})=39.31\pm 0.04+\log h 
\label{eq:L-density}.
\end{equation}
%
%, including the correction for the underdensity of the northern 
% equatorial stripe compared to the global value. 
We estimate the systematic error to be ${+0.10 \atop -0.07}$.
   % is 0.10 dex (38\%)
   % lower than the value by Gallego et al., which is $39.36\pm0.04$
   % taking their revised value and after small redshift corrections.  
   % We consider that this difference is dominantly caused by the 
   % neglect of stellar
   % absorption corrections in Gallego et al. 
   % As we discussed earier,
   % the neglect of stellar absorptions leads to a gross overestimate of
   % extinction corrections\footnote{
% We have examined the catalogue of Gallego et al. (1996). Their selective
% extinction $E(B-V)$ decreases clearly with the H$\alpha$ line width,
% which is the feature that occurs if stellar absorption is ignored.
% In fact their mean value of $E(B-V)\simeq 0.63$ ($A_V=2.0$) 
% is 1.8 times as large
% as the extinction we obtained. We confirmed that this happens when
% extinction corrections are estimated ignoring stellar absorption.
% This overcorrection of extinction leads to a 35\% overestimate of 
% H$\alpha$ luminosity.}.  We checked that the H$\alpha$ LF is
   % larger than the true value by 0.12 dex if we ignore the
   % corrections for stellar absorption. This is nearly the difference
   % we have seen between the two values: the remainder may be ascribed to
   % the difference in the threshold and the sample variance.

2. Early- ($1.5\leq T\leq 3$) and late-type 
   $(3.5\leq T\leq 5)$ spirals are the dominant contributors
   to the total H$\alpha$ LF. The H$\alpha$ LFs for these two types of galaxies
   are similar, and they determine the shape of the total LF.
   The faint end slopes are somewhat steeper than those for the
   $r'$ band LF.
   Spiral galaxies ($1.5\leq T\leq 5$) produce 83\% of the
   total  H$\alpha$ luminosity density.

3. The contribution of early type galaxies (E and S0) is only 9\% of the
   total luminosity density. In particular the luminosity density
   from E (8 galaxies) is only 6\% that of all early type (E to S0) galaxies.
   The characteristic luminosity of the H$\alpha$ LF for the E-S0 types
   is about 0.7 dex fainter than those for spiral galaxies.
   While our `best fit' to the H$\alpha$ LF indicates a decline 
   towards the faint end,
   we are not able to conclude whether the  H$\alpha$ 
   LF for early type galaxies actually
   declines because of the small number of 
   galaxies and the omission of very weak
   H$\alpha$ emitters. The inclusion of weak 
   emitters [EW(H$\alpha$)$\ge-0.8$({\AA})] lifts the faint end 
   slope, although $\alpha$ is still larger than $-1$ ($\alpha\approx -0.5$).

4. The H$\alpha$ LF for Im type galaxies shows a steep faint end slope. 
   This result is 
   similar to that found for the $r'$ band LF. Im galaxies 
   contribute 5\% of the local H$\alpha$ luminosity density. This is rather
   a significant fraction if we consider the fact that only 0.7\% of
   galaxies (10 galaxies) in our sample are Im type and their broad 
   band luminosity 
   is fainter
   than that of normal galaxies. We also note in Figure 1 that
   all Im galaxies are strong H$\alpha$ emitters. 

5. Most of the galaxies with unclassified morphology (6 out of 7 galaxies) 
   are also strong
   H$\alpha$ emitters [EW(H$\alpha)>20$\AA]. 
   These unclassified galaxies contain mergers.

\subsection{Comparison with Gallego et al.}

Since the work of Gallego et al. (1995) provides the current fiducial
value for the star formation rate at zero redshift,
we shall present a detailed comparison of their results and our
H$\alpha$ LF. 

The luminosity function and luminosity density we derived are consistent
with the findings of Gallego et al. who derived $\alpha=-1.3\pm0.2$,
and $\log L^*=41.56\pm 0.08-2\log h$. While this $L^*$ looks somewhat smaller
than our value, $L^*$ and $\alpha$ are correlated such that 
$\Delta (\log L)\approx -0.4\Delta \alpha$; so if we adjust $\alpha=-1.43$,
then we have $\log L^*=41.61\pm 0.08-2\log h$ which is smaller than our 
$\log L^*=41.68\pm0.10-2\log h$
only by 1 $\sigma$. 
The luminosity density 
$\log {\cal L}({\rm H}\alpha)=39.38\pm 0.04+\log h$
is larger than  (\ref{eq:L-density}) by about 20\% (0.07 dex).
We consider, however, that these rather good agreements are fortuitous.

Gallego et al. did not take account of stellar absorption. 
%As we discussed earier, 
The neglect of stellar absorptions leads to 
a substantial overestimate of extinction corrections (see section 
\ref{sec:extinc} above).
We find that their selective extinction $E(B-V)$ tabulated in the
catalogue of Gallego et al. (1996) show a conspicuous decrease as 
the H$\alpha$ equivalent width increases, which is the feature that occurs if 
stellar absorption is ignored.
In fact their mean value of $E(B-V)\simeq 0.64$ ($A_V=2.05$) is 1.9 
times as large as the extinction we obtained. This overestimate
of the absorption makes the luminosity
density larger by 0.16 dex than the true value (see Table 3).
The inclusion of AGN also leads to an overestimate of the luminosity
density by $<0.07$ dex. 

Gallego et al. set the threshold EW(H$\alpha$+[NII])$>$10\AA (effective
H$\alpha$ EW 7\AA); this selection would cause a 10\% decrease in the
H$\alpha$ luminosity density. What gives a strong effect is their
sample incompleteness, which compensates the significant overdensity
seen above. The surface
density of H$\alpha$ emitters with EW(H$\alpha$+[NII])$>$10\AA~ is
176/471=0.37 per sq. deg.  This is compared with 306/230=1.33 per sq. deg.
when we select emission line galaxies from the SDSS sample in a manner
that satisfies the same EW threshold
and $z<0.045$ as in Gallego et al. 
The EW distributions of the two samples are compared in
Figure 10, which shows where Gallego et al.'s sample is incomplete. Note
that we have applied aperture corrections for the EW distribution of the
SDSS sample so that the line flux represents that from the entire galaxy.
%as would be expected in the objective prism surveys. 
It appears that the
underestimate due to sample incompleteness 
cancels the overestimates from the neglects of
stellar absorptions and contamination from AGN.

\section{Correlation of H$\alpha$ emissivity with galaxy properties}

\subsection{H$\alpha$ emissivity and galaxy morphology}

Figure 1 and Table 1 above show the clear trend that the fraction of
H$\alpha$ emission galaxies increases as 
$T$ moves from early to late types. If we set the threshold
of EW$>1$\AA, the fraction of galaxies with H$\alpha$ emission is
5\% (4\% if we remove galaxies with AGN activity) for E galaxies, 
15\% (12\%) for S0 galaxies,
54\% (44\%) for S0/a-Sb and 86\% (80\%) for Sbc-Sd galaxies.
We have seen that S0/a-Sb and Sbc-Sd contribute
nearly the same amount to the luminosity density. This is due to a larger
number of S0/a-Sb type galaxies than that of Sbc-Sd (by a factor
of 1.5),
whereas the fraction of emitters in the former class is less by
1.6 times than the latter; these changes in the two factors 
cancel and leave the luminosity density unchanged.
In our sample, the fraction of H$\alpha$ emitters is 100\% for Im galaxies.
A high fraction ($\approx 86$\%) is also recorded for galaxies of
unclassified types. Conversely, 89\% of emitters are spiral galaxies,
and they contribute to 83\% of the H$\alpha$ luminosity density. 
Note that elliptical galaxies
are not necessarily inactive (Fukugita et al. 2003).

Galaxies in each morphological class contain strong H$\alpha$ emitters. If
one makes a cut at high EW threshold, say at 10\AA, the average of
EW for those emitters look nearly independent of morphology. The 
most important difference that varies with morphology is 
in the increasing number
of non-emission or weak emission galaxies towards early types. 
Therefore the average EW decreases to early types when the average 
is taken over all
galaxies including non-emission galaxies.
The `average' emission equivalent width depends 
on how to set the threshold for emission galaxies.

Kennicutt and Kent (1983) showed a clear correlation of 
EW(H$\alpha$+[NII]) with morphology, notably the absence of
strong emitters in earlier types (Sa or earlier).  We consider 
this absence of emitters being due to their small sample, and in particular
to the fact that bona fide early-type galaxies are selected 
before the observation.

\subsection{H$\alpha$ emissivity and galaxy colours}

The natural assumption is that H$\alpha$ emission galaxies, being 
hosts to vigorous star formation, should have bluer colours than
galaxies that are bereft of H$\alpha$ emission.
Tresse \& Maddox (1998) claimed, however, that there is no correlation 
between H$\alpha$ luminosity $L({\rm H}\alpha)$ and $(V-I)$ colour. 
We also find in our
sample that the correlation between H$\alpha$ luminosity and $g^*-r^*$
colour is weak (Figure 11a). Also, we do not see correlations with other
colour indices, e.g., $u^*-g^*$.
The correlation we should actually expect, however, is not 
between total H$\alpha$
luminosity and colour, but between  H$\alpha$ emissivity per
unit mass (or per broad band luminosity) and the colour.
To see this point we plot in Figure 11b the correlation of the H$\alpha$
luminosity divided by $i'$ band luminosity with 
$g^{*}_{P}-r^{*}_{P}$ colour. The expected correlation is observed:
there is a 2 mag change in H$\alpha$ luminosity normalised by
$i'$ band luminosity across 0.5 mag change in $g^{*}-r^{*}$ colour,
though the scatter is significant, especially for strong H$\alpha$
emitters.

\section{Implications of our results}

We have derived the H$\alpha$ luminosity function for the nearby universe
and have made a break down into morphological types of galaxies. Our sample
contain 665 emission line galaxies, which are compared, for example, to 
176 galaxies in Gallego et al. (1995), 159 galaxies in Sullivan et al.
(2000), and 110 galaxies in Tresse \& Maddox (1998); see Table 4, which
gives a comparison of the present result with those of the earlier work
made for low redshifts. 
Our observed range of L(H$\alpha$) is 0.4 dex deeper than Gallego et al., 
although it is shallower by 0.4-0.8 dex than surveys of
distant galaxies by Sullivan et al. (2000) and Tresse \& Maddox
(1998). 

The local H$\alpha$ luminosity density we have obtained is  
$10^{39.31\pm 0.04{+0.10\atop-0.07}}h$ erg s$^{-1}$Mpc$^{-3}$. 
The central value is lower than that of Gallego et al. (1995) but 
only by 0.08 dex, although this agreement is fortuitous. 
%So we do not need to modify the arguments in the literature, which were
%based on Gallego et al.

The significant discrepancy between Tresse \& Maddox (1998)  and 
Sullivan et al. (2000) for analyses at non-zero redshift
hinders us from drawing a definitive
conclusions as to the evolution of H$\alpha$ emissivity.
In comparing the characteristic luminosities we note that 
$L^*$ and $\alpha$ are strongly correlated. Assuming
an empirical law, $d\log L^*/d\alpha\approx -0.4$, Tresse \& Maddox's
value is ($41.64\pm0.13-2\log h$) dex and Sullivan et al.'s is ($42.03\pm0.14-2\log h)$ dex
if we adjust $\alpha$ to our $-1.43$.
Our value is in between the two. Our faint end slope is also between
the two groups of authors.  On the other hand, the H$\alpha$ luminosity
density of the two  groups  
($10^{39.66\pm0.04}$, $10^{39.43\pm0.06}$)$h$ erg s$^{-1}$Mpc$^{-3}$ 
are larger than
our value by 2.2$-$1.3 times.
We emphasize, however, that the samples at non-zero redshifts are too
small, and the uncertainty in the correction of 
the spectroscopically obtained flux
to total flux might also be significant; the definitive conclusion
may not be drawn as to the evolution at low redshift. The evolution
can be concluded, however, if we compare our results with those
at high redshifts (e.g., Glazebrook et al. 1999; Tresse et al. 2002).

The conversion of H$\alpha$ luminosity density to the global star 
formation rate per unit volume depends further on models of star
formation and absorption of the ionising flux in the star-forming
regions (Charlot \& Longhetti 2001).
If we adopt the conversion factor of 
Glazebrook et al. (1999) for the case BC96(kl96) $Z/Z_\odot=1.0$, and
Salpeter IMF (with the lower mass cutoff at 0.1$M_\odot$): 
$L({\rm H}\alpha)=1.35\times 10^{41}$erg s$^{-1}$
for star formation rate of 1 $M_\odot$ yr$^{-1}$, we obtain
the global star formation rate $\psi$: % $M_\odot$ yr$^{-1}$Mpc$^{-3}$:
\begin{equation}
\psi\simeq 0.015 h M_\odot{\rm yr}^{-1}{\rm Mpc}^{-3},
\end{equation}
ignoring internal absorption of the ionising flux.

Some early type galaxies, even elliptical galaxies, emit strong 
H$\alpha$ by star formation activity, almost as strong as that of 
late spiral galaxies.
This contradicts the conventional wisdom that all ellipticals formed in
the early universe. The fraction of H$\alpha$ emitters, however, 
is quite small: it is only 1\%
if we take a threshold of EW(H$\alpha)>$10\AA~(and 5\% with 
EW(H$\alpha)>1$\AA): the majority of elliptical galaxies no longer have
significant star formation.
On the other hand, there are many late spiral
galaxies that do not show H$\alpha$ emission (in our sample all
Im galaxies are H$\alpha$ emitters). The presence
of non star-forming spiral galaxies is consistent with the
view that star formation in disc galaxies is intermittent,
but the fraction of star forming galaxies with $>1M_\odot$yr$^{-1}$
($\approx 50$\%) indicates that in spiral galaxies such
star formation activity takes place about half the time. 
The quantity that shows a strong correlation with
morphology of galaxies is the fraction of H$\alpha$ 
emitters, or the mean H$\alpha$ luminosities, not the 
H$\alpha$ luminosities of individual galaxies.

\vspace{10pt}

\noindent
{\bf Acknowledgements}

We should like to thank Stuart Ryder for
his useful communications and Christy Tremonti for useful
discussions and comments to the preliminary manuscript.
Funding for the creation and distribution of the SDSS Archive has been provided
by the Alfred P. Sloan Foundation, the Participating Institutions, the National
Aeronautics and Space Administration, the National Science Foundation, the US
Department of Energy, the Japanese Monbukagakusho, and the
Max-Planck-Gesellschaft. The SDSS Web site is http://www.sdss.org/. 
The SDSS is managed by the Astrophysical Research Consortium (ARC) for the
Participating Institutions. The Participating Institutions are The University of
Chicago, Fermilab, the Institute for Advanced Study, the Japan Participation
Group, Johns Hopkins University, Los Alamos National Laboratory, the
Max-Planck-Institut f\"{u}r Astronomie, the Max-Planck-Institut
f\"{u}r Astrophysik, New Mexico State
University, the University of Pittsburgh, Princeton University, the US
Naval Observatory, and the University of Washington.
MF is supported in part by the Grant in Aid of the Ministry of Education.

\clearpage

%table 1
\begin{table*}
\begin{center}
\caption{Number of galaxies in our sample}
\begin{tabular}{lccccc}
\tableline \tableline
  & T=$-1\sim6$ & T=$0\sim1$ & T=$1.5\sim3$ & T=$3.5\sim5$ & T=$5.5\sim6$ \\
\tableline
Total                      & 1482 &  597 &  518 &  350 &   10 \\
H$\alpha$ detected         &  665 &   67 &  282 &  300 &   10 \\
AGNs by Eq.(\ref{eq:agn2}) &   89 &   13 &   54 &   21 &    0 \\
\tableline
\end{tabular} 
\end{center}
\end{table*}%

%table 2
\begin{table*}
\begin{center}
\small
\caption{Luminosity Function Parameters}
\begin{tabular}{rccccc}
\tableline \tableline
            & $\log L^{\ast}$
             &
              & $\phi^{\ast}$
               & \multicolumn{2}{c}{
                  $\log \cal L$(H$\alpha$)
                 ($h$ erg s$^{-1}$ Mpc$^{-3}$)} \\
                  \cline{5-6}
Morphology  & ($h^{-2}$ erg s$^{-1}$)
             & $\alpha$
              & ($0.001 h^{3}$Mpc$^{-3}$)
               & Integrated
                & Sum                           \\
\tableline
% T                  L*              alpha             phi*           lum-den        sum
 $-1\le T\le 6$ & $41.68\pm0.10$ & $-1.43\pm0.10$ & $2.78\pm1.13$  & $39.31\pm0.04$ & $39.33$ \\
$  0\le T\le 1$ & $41.02\pm0.14$ & $+0.79\pm0.77$ & $0.78\pm0.34$  & $38.13\pm0.22$ & $38.24$ \\
$1.5\le T\le 3$ & $41.70\pm0.13$ & $-1.40\pm0.15$ & $1.03\pm0.43$  & $38.88\pm0.03$ & $38.91$ \\
$3.5\le T\le 5$ & $41.71\pm0.21$ & $-1.53\pm0.21$ & $0.96\pm0.50$  & $38.96\pm0.08$ & $39.01$ \\
$5.5\le T\le 6$ & $\sim42.76$    & $\sim-1.77$    & $\sim0.006$    & $\sim38.13$    & $37.98$ \\
$T=-1$    &    &    &    &    & 37.80 \\
% error for L*, alpha  = error circle
% error for phi*, l.d. = jack-knife with 1 sigma
\tableline
\end{tabular} 
\end{center}
\end{table*}%

%table 3
\begin{table*}
\begin{center}
\caption{Summary of Systematic Errors}
\begin{tabular}{lcc}
\tableline \tableline
Item     &      & Change in $\cal L$(H$\alpha$) \\
     &  & (factor)                      \\
\tableline
EW(H$\alpha$) threshold  ($1.0$\AA)                      &       \\
\hspace{0.5cm} decreased to $-0.8${\AA}           & *                & 1.04  \\
\hspace{0.5cm} increased to $7${\AA}             &                & 0.90  \\
%\hspace{0.5cm} $= 20${\AA}           &                 & 0.59  \\
%\hspace{0.5cm} $= 13${\AA}           &                 & 0.75  \\
%\hspace{0.5cm} $= 3${\AA}            &                 & 0.97  \\
%\hspace{0.5cm} $= 0${\AA}            &                 & 1.02  \\
Stellar absorption correction  $\pm 0.5$\AA~    &  *         & 0.97$-$1.07\\
%\hspace{0.5cm} EW$_{abs}$(H$\alpha)=0${\AA}  &         & 0.90  \\
%\hspace{0.5cm} EW$_{abs}$(H$\alpha)=3${\AA}  &         & 1.18  \\
\hspace{0.5cm} entirely neglected              &          & 1.46  \\
Extinction law   (Seaton 1979)                  &       \\
\hspace{0.5cm}  O'Donnell (1994), $R_v=3.1\pm0.2$   & *  & 0.98$-$1.10  \\
Errors in the estimate of $A_V=1.03$       &    *     & 0.90$-$1.10  \\
Remove weak H$\beta$ 'AGN's                    &   *    & 0.95  \\
Include all AGNs                               &        & $<$1.18  \\
Nebulae Temperature ~ 6000$-$15000 K            &  *    & 0.97$-$1.03  \\
Aperture corrections:  \\
   \hspace{0.5cm} RD94 profile with $\pm1\sigma$  &  *  & 0.91$-$1.19  \\
\hspace{0.5cm} scale with the broad band flux&         & 0.75$-$0.80  \\
\tableline
Total systematics                     &     &  0.85$-$1.25  \\
\tableline
\end{tabular}
\end{center}
*) Counted as systematic errors, and added in quadrature to estimate the
   total systematic errors.

\end{table*}

%table 4
\begin{table*}
\begin{center}
\caption{Comparison with the earlier work ($h=1$)}
\begin{tabular}{lrrrr}
\tableline \tableline
               & Gallego 95 & Tresse-Maddox 98 & Sullivan 00 & {\bf This Work}\\
\tableline
Survey area        & 471.4 deg$^2$        & 500 arcmin$^2$    & $\sim10$ deg$^2$   & 229.7 deg$^2$\\
Mean redshift      & $\sim 0.025$         & $0.21$          & $0.15$             & $0.054$\\
%Redshift range     & $0 \le z \le 0.045$  & $0 \le z \le 0.3$ & $0 \le z \le 0.4$  & $0.01 \le z \le 0.12$ \\
Size of the sample  & 176            & 110               & 159                & 665\\
$\log L^{*}$   & $41.56\pm0.08$          & $41.61\pm0.13$    & $42.11\pm0.14$     & $41.68\pm0.10$ \\
$\alpha$       & $-1.3\pm0.2$             & $-1.35\pm0.06$    & $-1.62\pm0.10$     & $-1.43\pm0.10$ \\
$\log\phi^{*}$ & $-2.3\pm0.2$             & $-2.09\pm0.09$    & $-3.04\pm0.20$     & $-2.56\pm0.30$ \\
$\log \cal L$(H$\alpha$) & $39.38\pm0.04$ & $39.66\pm0.04$    & $39.43\pm0.06$     & $39.31\pm0.04$ \\
\tableline
\end{tabular}
\end{center}
\end{table*}

%figure 1
%%\vspace{0.5cm}
\begin{figure}
%%\psbox[width=8.5cm,aspect=1.0]{fig_rewha.eps}
%\plotone{fig_rewha.eps}
\plotone{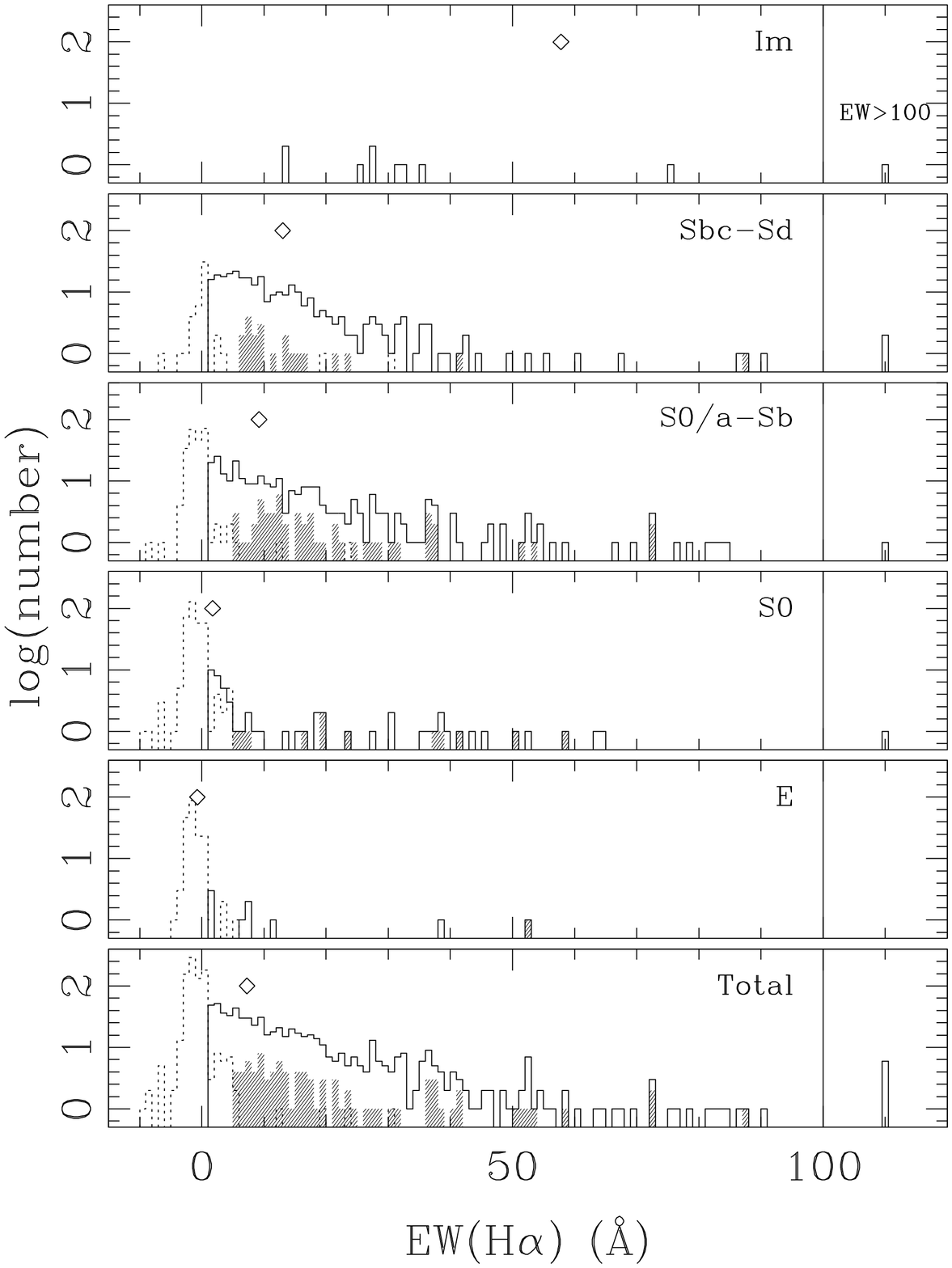}
\caption{
Distribution of rest EW(H$\alpha$) for each morphological type
       of galaxies with the rest EW(H$\alpha$)$>1$\AA~and the detection
       at $>2.5\sigma$. The lowest panel shows the total sample. Those
       plotted near the right margin are numbers of galaxies with rest 
       EW(H$\alpha$)$>100$\AA. The galaxies that satisfy the AGN
       criterion are denoted by shading. Dotted histogram show
       galaxies with either rest EW(H$\alpha$)$<1$\AA~or
       the emission detection below 2.5$\sigma$.
       The diamonds show the
       mean EW(H$\alpha$).}
\end{figure}%

%figure 2
%%\vspace{0.5cm}
\begin{figure}
%%\psbox[width=8.5cm,aspect=1.0]{fig_hahb.eps}
\plotone{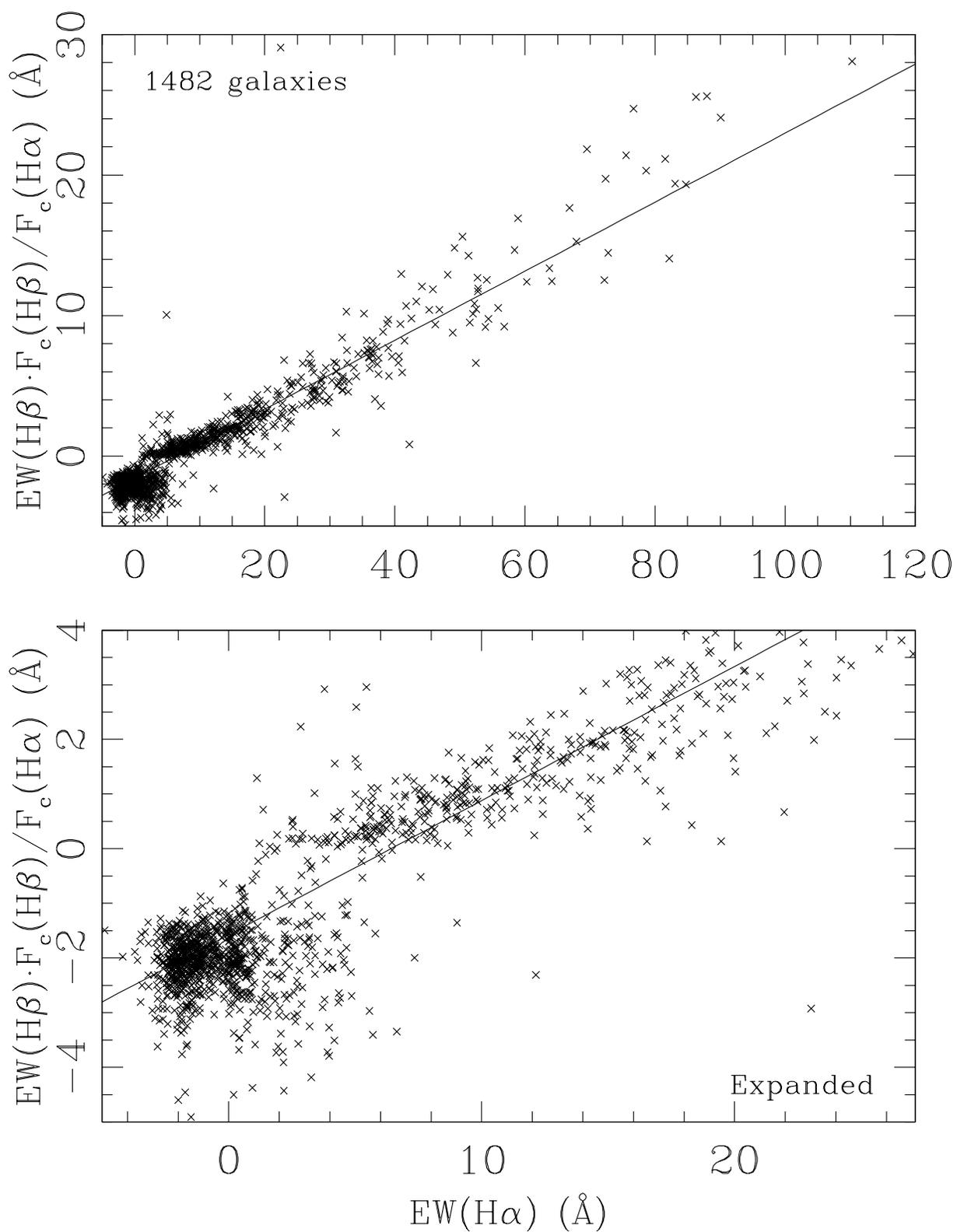}
%\plotone{fig_hahb.eps}
\caption{Relation between the rest EW(H$\alpha$) and 
       rest EW(H$\beta$)$\times [F_c$(H$\beta)/F_c$(H$\alpha$)],
       where $F_c$ is the flux density of the continuum. The region
       of small EW is expanded in the lower panel. The lines show the fit
       of eq. (1).}
\end{figure}%

%figure 3
%%\vspace{0.5cm}
\begin{figure}
%%\psbox[width=8.5cm,aspect=1.0]{fig_hahg.eps}
\plotone{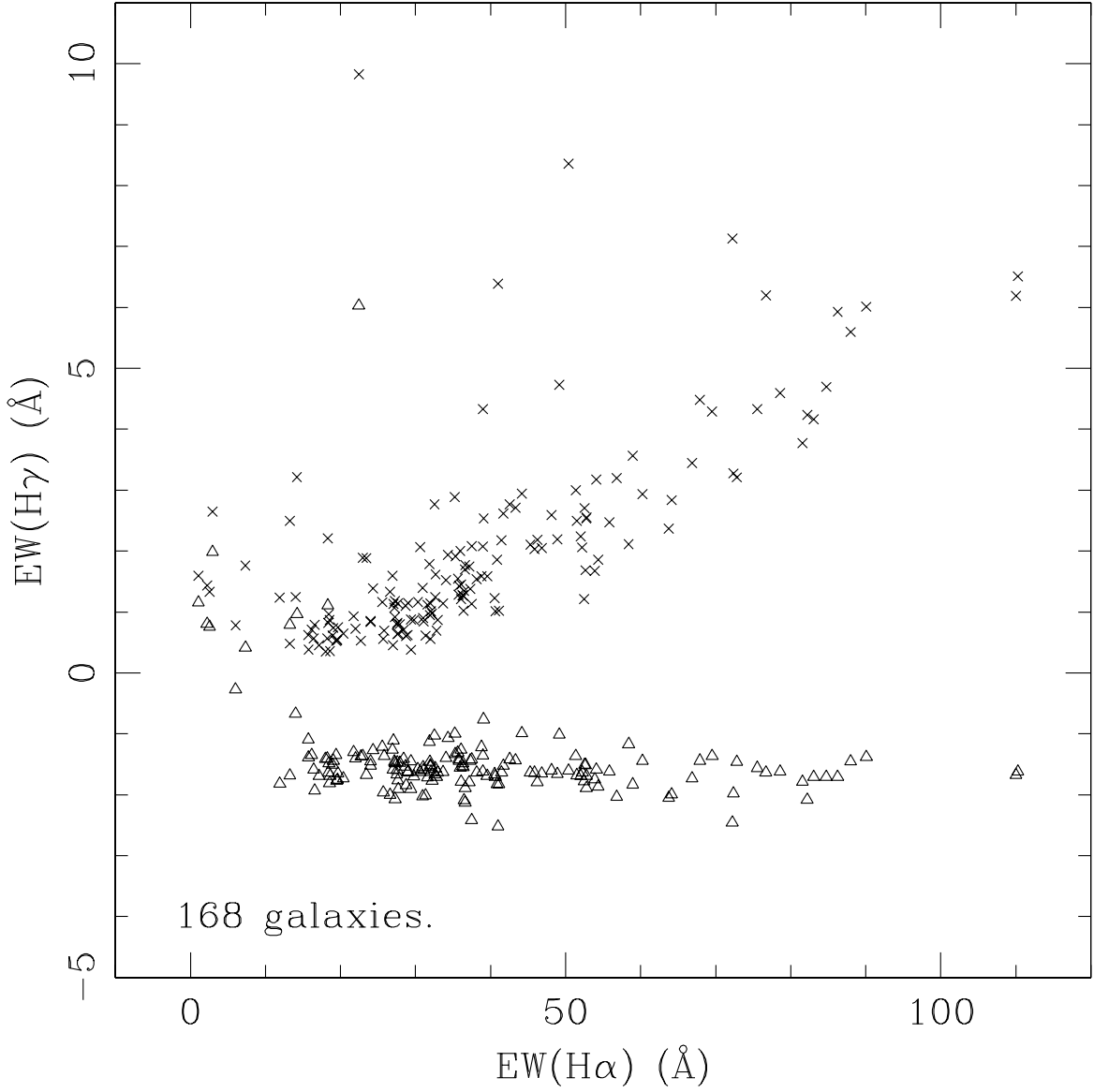}
%\plotone{fig_hahg.eps}
\caption{H$\gamma$ equivalent widths (crosses), and stellar absorption
       equivalent widths of H$\gamma$ (triangles) plotted as a function 
       of H$\alpha$ emission equivalent widths.}
\end{figure}%

%figure 4
%%\vspace{0.5cm}
\begin{figure}
%%\psbox[width=8.5cm,aspect=1.0]{fig_ext.eps}
\plotone{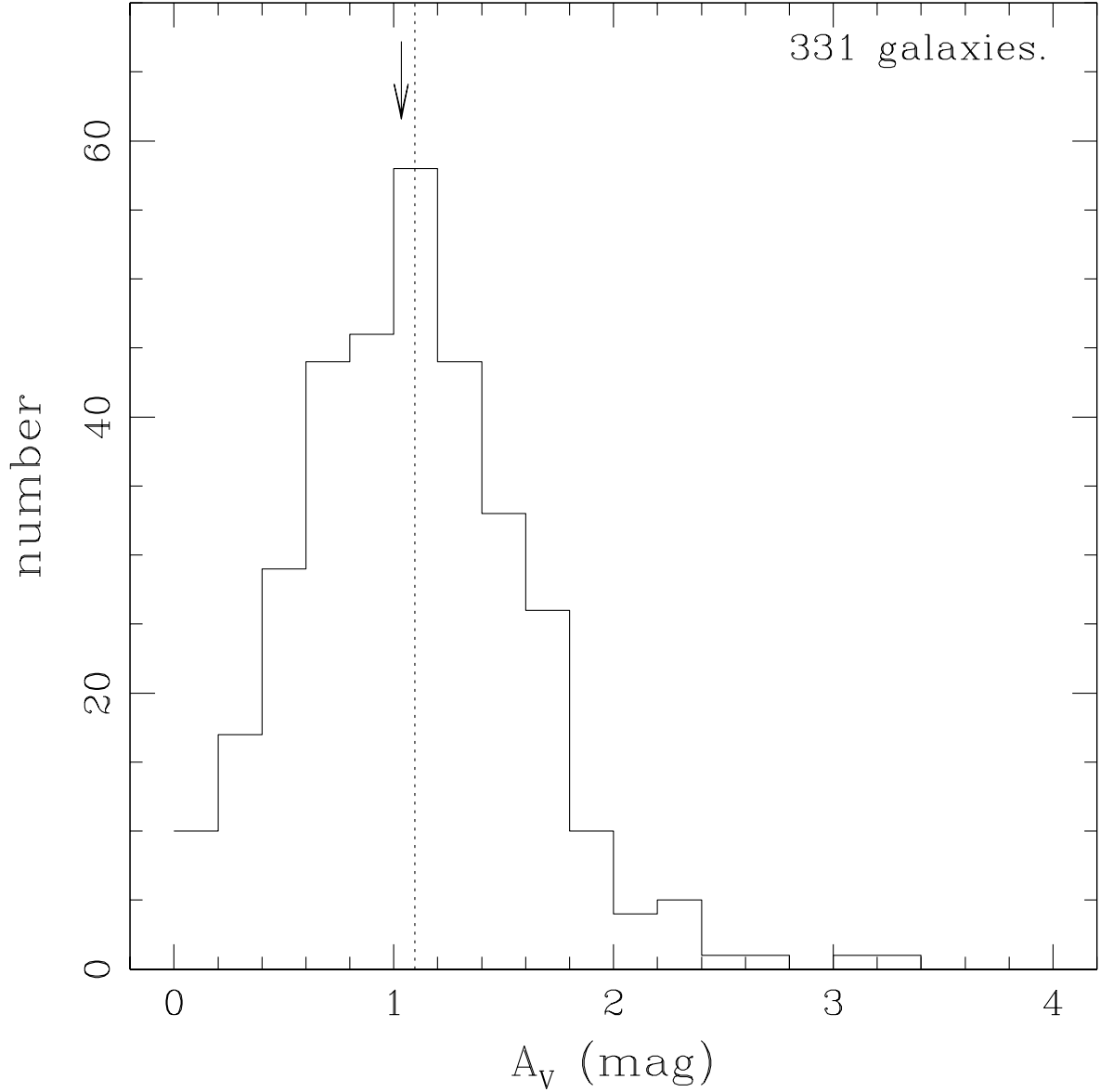}
%\plotone{fig_ext.eps}
\caption{Distribution of extinction estimated from Balmer line ratios for
       331 galaxies with strong emission (see text).
       The line shows the mean. The arrow indicates the value derived from
       eq. (1).}
\end{figure}%

%figure 5
%%\vspace{0.5cm}
\begin{figure}
%%\psbox[width=8.5cm,aspect=1.0]{fig_RD94_1.eps}
\plotone{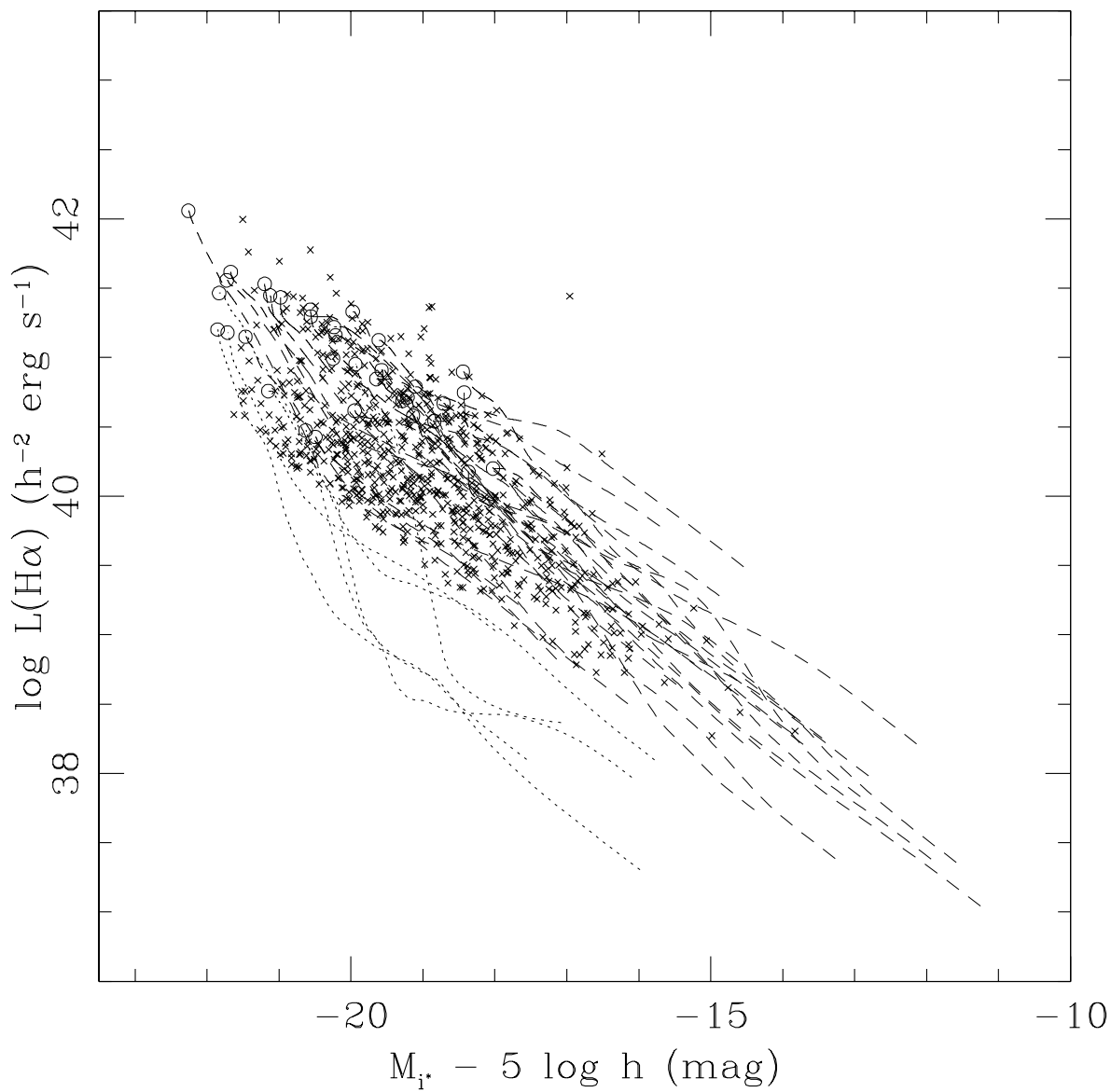}
%\plotone{fig_RD94_1.eps}
\caption{Relation between the H$\alpha$ flux and the $i'$ band flux. The
       dashed lines show galaxies measured by RD94 at varying
       apertures (the point moves left/upwards as the aperture size increases).
       The circles indicate the `total flux' given by RD94.
       The crosses are SDSS spectrophotometric data for the 3$''$ aperture,
       calibrated by photometrically measured fluxes. The dotted lines
       show 6 profiles that are outliers.}
\end{figure}%

%figure 6
%%\vspace{0.5cm}
\begin{figure}
%%\psbox[width=8.5cm,aspect=1.0]{fig_RD94_2.eps}
\plotone{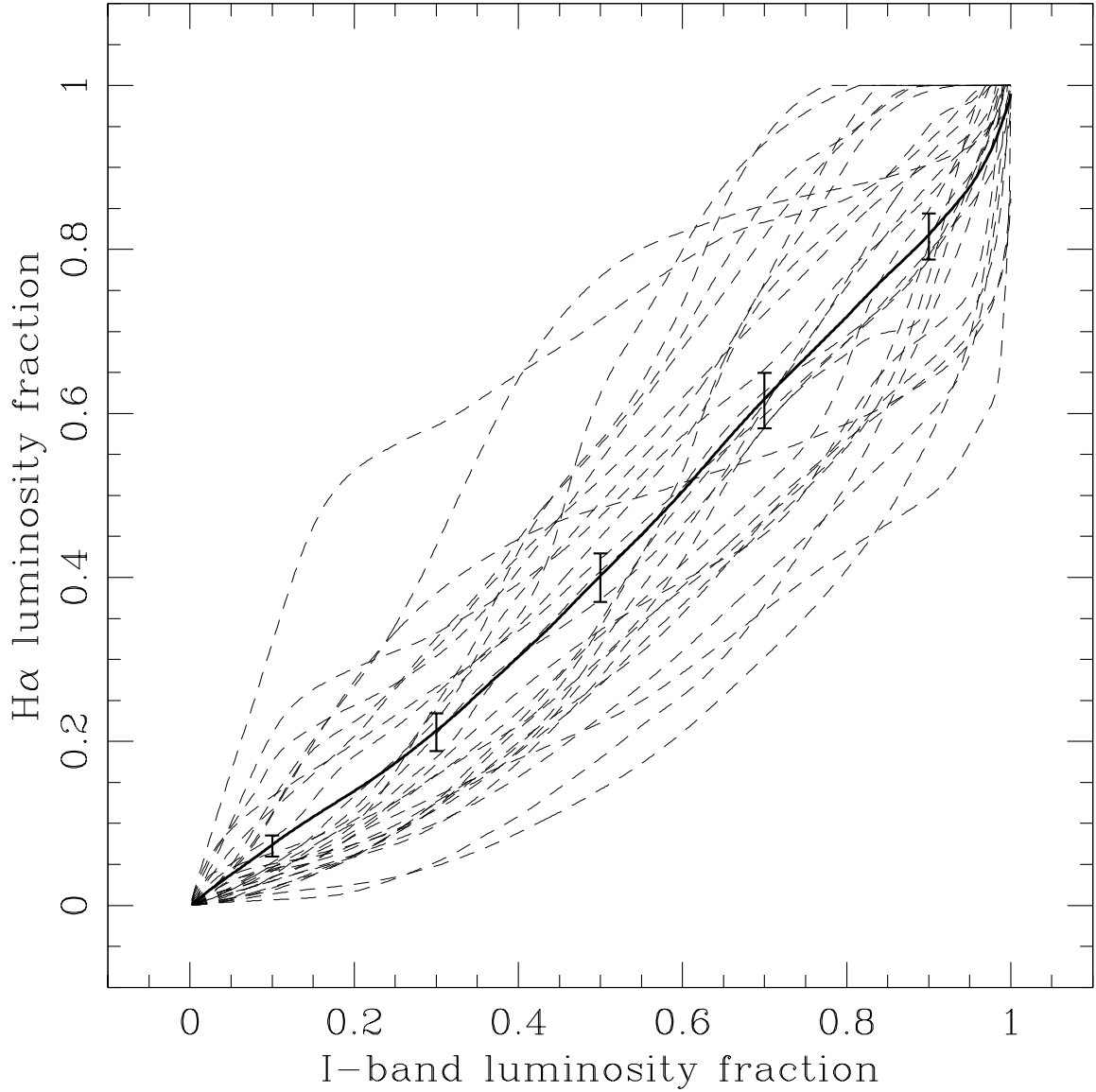}
%\plotone{fig_RD94_2.eps}
\caption{Growth curve of the H$\alpha$ flux against that of
       the $I$ band flux. The thick line is the composite
       growth curve, and error bars are one sigma of the mean
       at several representative points.
       The thin dashed curves are those
       of individual galaxies in RD94. }
\end{figure}%

%figure 7
%%\vspace{0.5cm}
\begin{figure}
%%\psbox[width=8.5cm,aspect=1.0]{fig_LHa2.eps}
\plotone{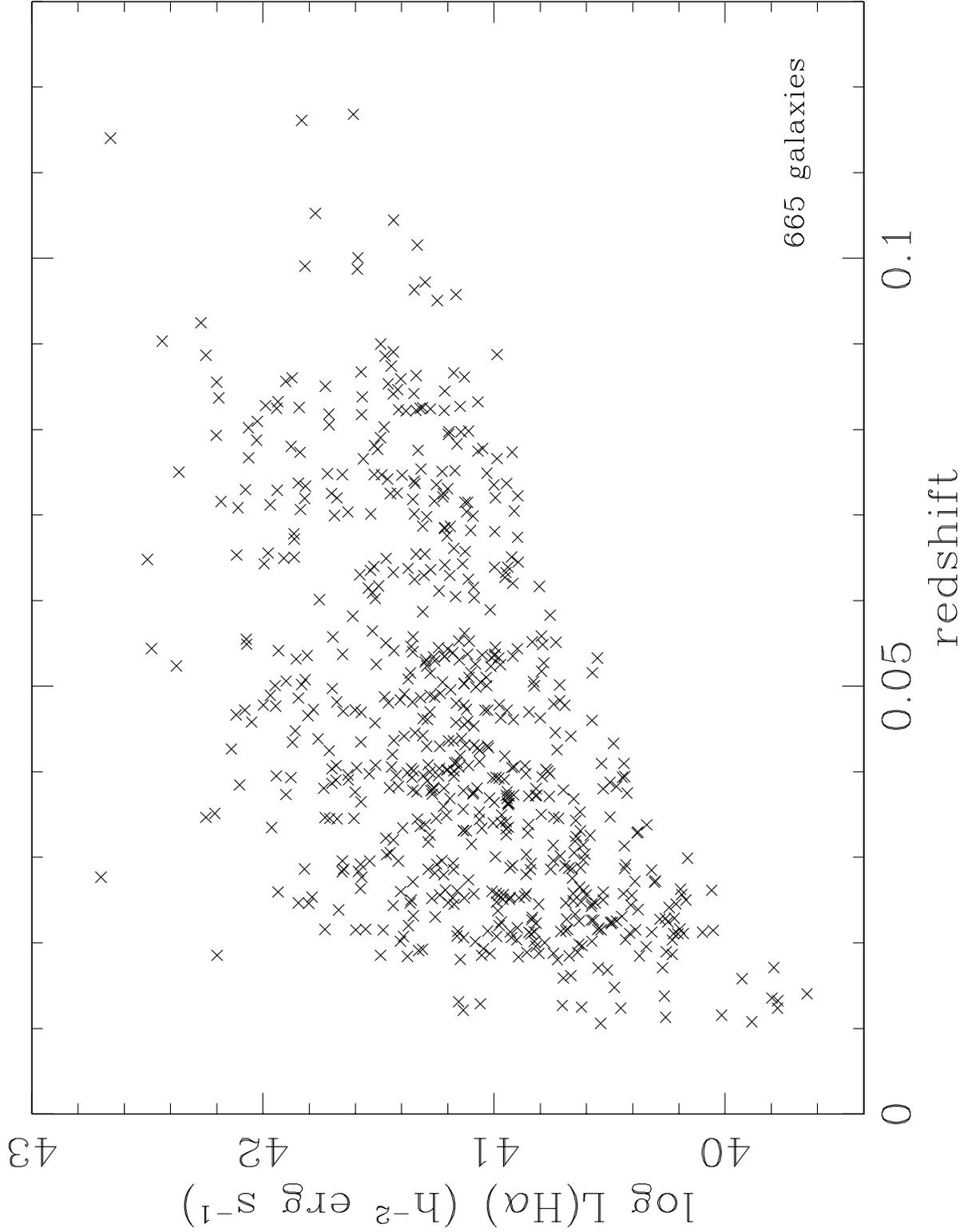}
%\plotone{fig_LHa2.eps}
\caption{H$\alpha$ luminosity as a function of redshift after the aperture
       correction.}
\end{figure}%

%figure 8
%%\vspace{0.5cm}
\begin{figure}
%%\psbox[width=8.5cm,aspect=1.0]{fig_half.eps}
\plotone{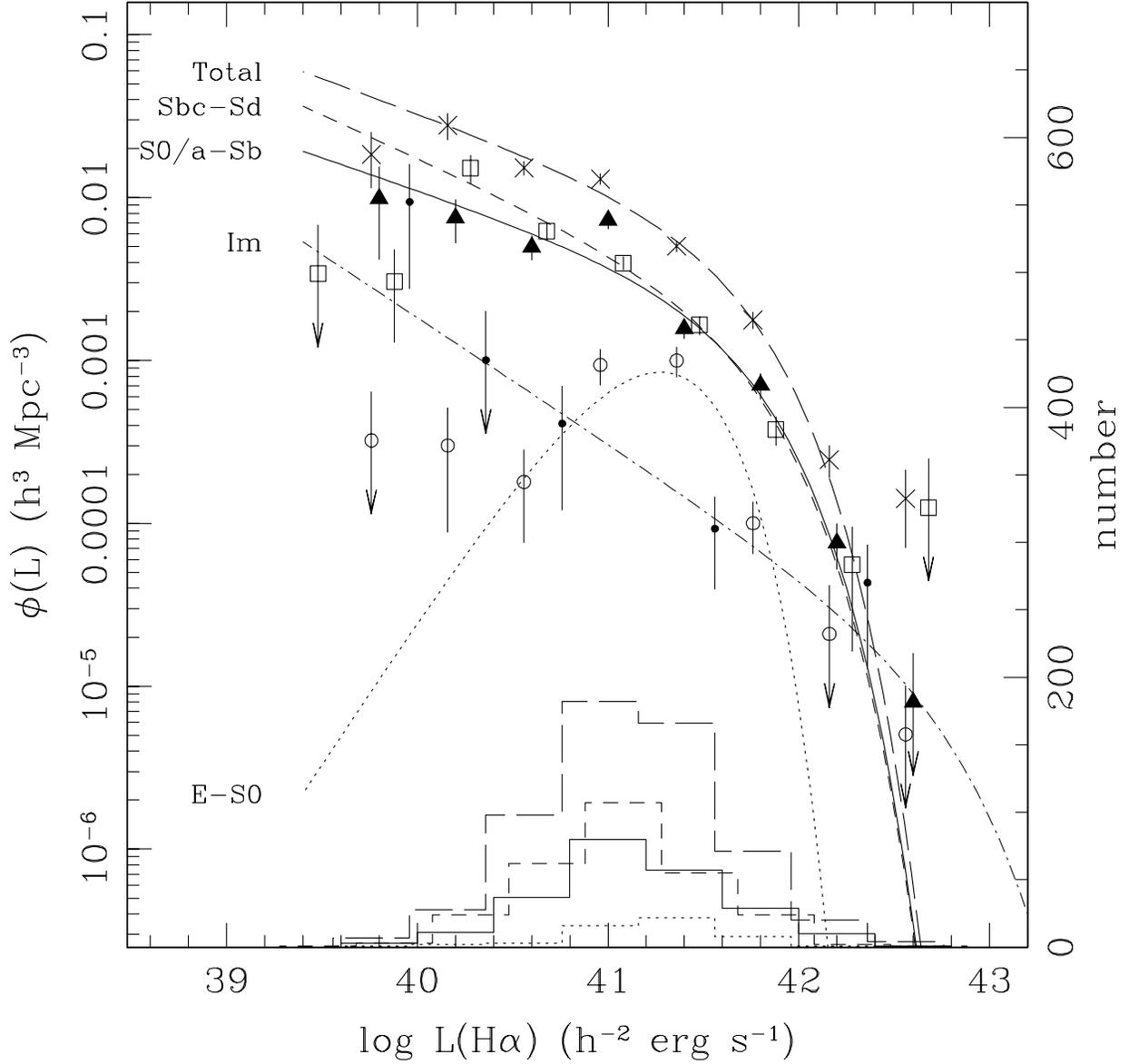}
%\plotone{fig_half.eps}
\caption{H$\alpha$ LF for each morphological type of galaxy. The data
       points are step wise LF, and the curves show Schechter function
       fits.  The histograms are actual number of galaxies used to
       derive the LF. Data symbols are: total (cross), E-S0 (open circle),
       S0/a-Sb (solid triangle), Sbc-Sd (open square), Im (solid circle).}
\end{figure}%

%figure 9
%%\vspace{0.5cm}
\begin{figure}
%%\psbox[width=8.5cm,aspect=1.0]{fig_err.eps}
\plotone{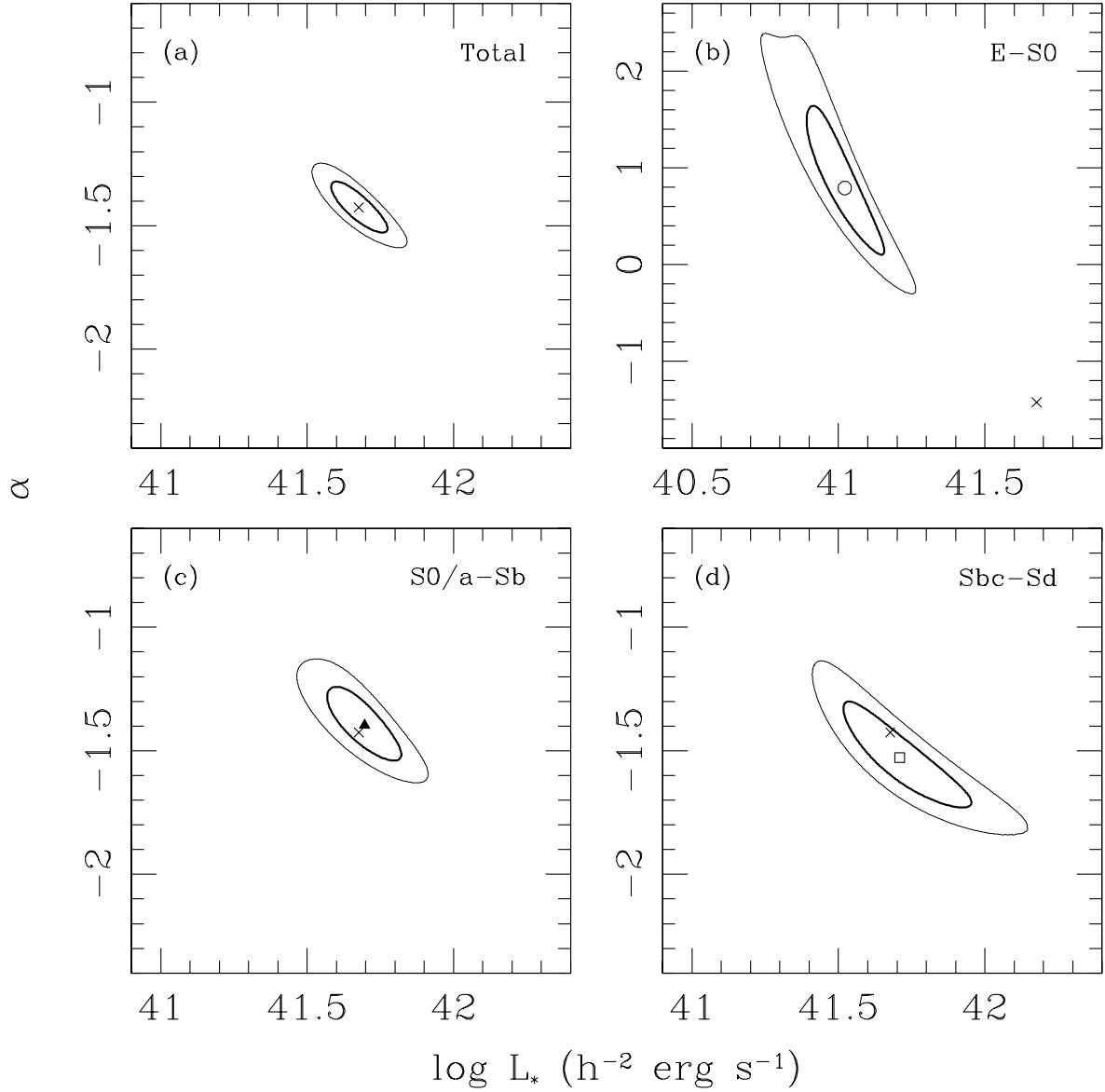}
%\plotone{fig_err.eps}
\caption{Schechter function parameters of H$\alpha$ LF: (a) total sample,
       (b) E-S0, (c) S0/a-Sb, and (d) Sbc-Sd.  The two contours show
       1 $\sigma$ and 2 $\sigma$ errors. The cross in each panel shows 
       the position of the parameter for the total sample.}
\end{figure}%

%figure 10
%%\vspace{0.5cm}
\begin{figure}
%%\psbox[width=8.5cm,aspect=1.0]{fig_ew_g95.eps}
\plotone{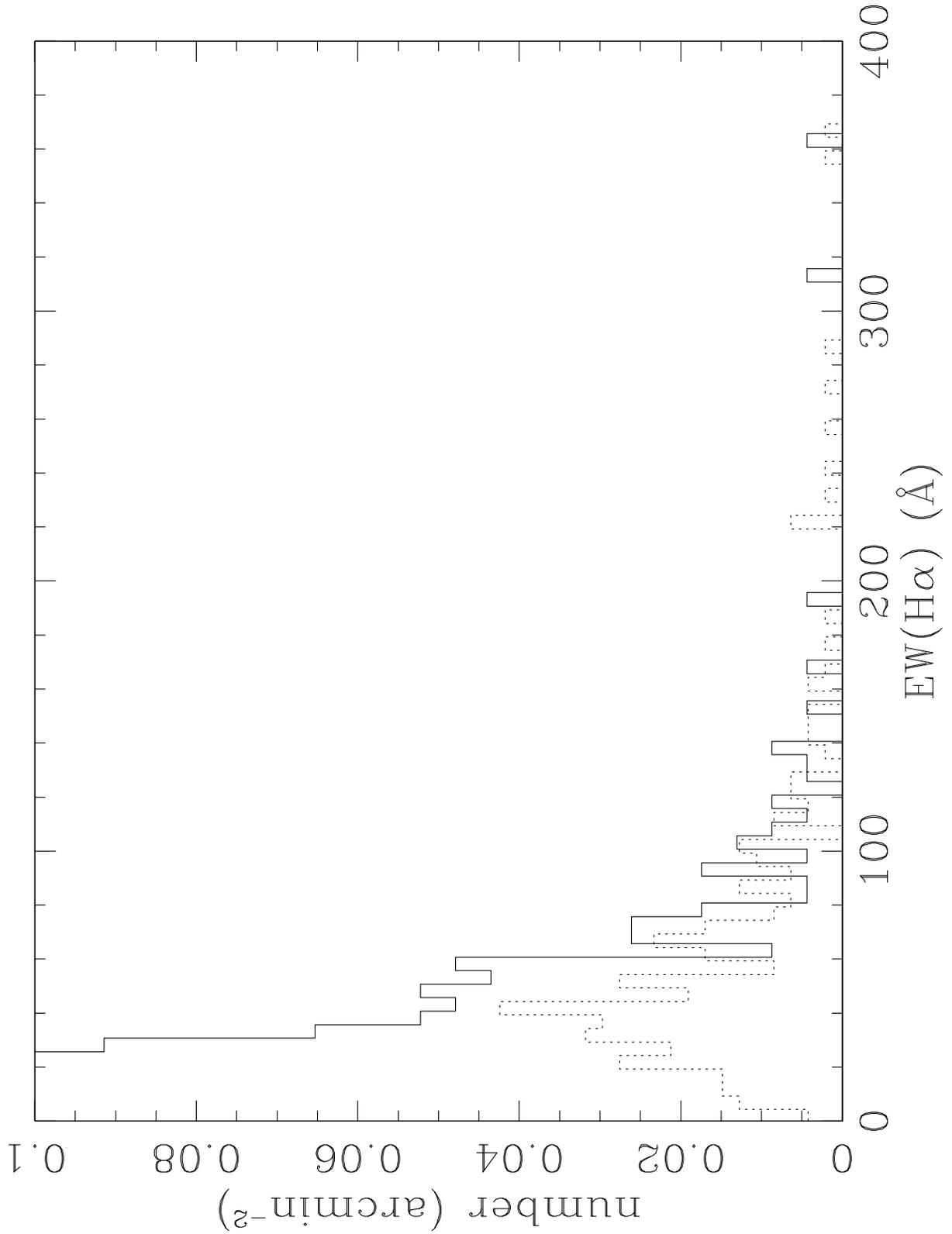}
%\plotone{fig_ew_g95.eps}
\caption{Comparison of the rest EW distribution of Gallego et al. (1995)'s
       sample (dotted lines) with ours (solid lines).}
\end{figure}%

%figure 11
%%\vspace{0.5cm}
\begin{figure}
%%\psbox[width=8.5cm,aspect=1.0]{fig_col_LHa.eps}
\plotone{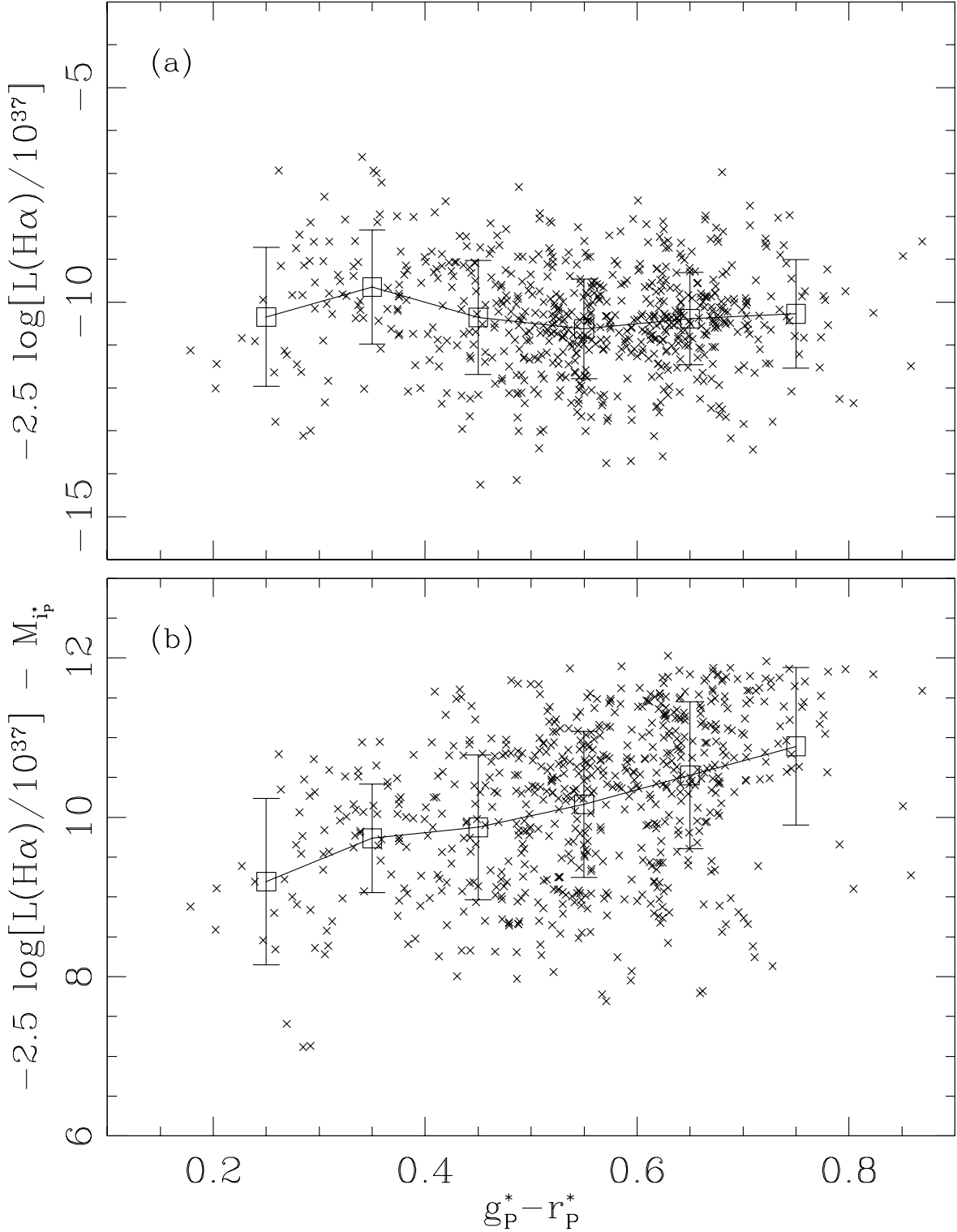}
%\plotone{fig_col_LHa.eps}
\caption{(a) H$\alpha$ luminosity (in magnitude scale) 
       plotted as a function of $g^*-r^*$ 
       colour (in Petrosian magnitudes). (b) H$\alpha$ luminosity
       normalised by $i'$ band luminosity plotted as a function of $g^*-r^*$ 
       colour. The squares are the mean of the data and the error bars
       stand for rms.}
\end{figure}%

\end{document}